\documentclass{llncs}

\usepackage{amsmath}
\usepackage{amsfonts}
\usepackage{changebar}
\usepackage{ifthen}
\usepackage{multirow}
\usepackage{graphicx,amssymb}
\usepackage{latexsym}
\usepackage{xspace}
\usepackage{url}
\usepackage{times}
\usepackage{enumerate}
\usepackage{epsfig}
\usepackage{empheq}
\usepackage{qtree}
\usepackage{dsfont}
\usepackage[usenames,dvipsnames]{xcolor}
\usepackage{tikz}
\usetikzlibrary{arrows,shapes}%
\usepackage{tkz-graph}

\newcommand{\grap}[2]{\ensuremath{\langle #1,#2 \rangle}}

\newcommand{\expa}{\ensuremath{\mathit{exp}}}
\newcommand{\unexpa}{\ensuremath{\mathit{unexp}}}

\newcommand{\ct}{\ensuremath{\mathsc{ct}}}
\newcommand{\st}{\ensuremath{\mathsc{st}}}

\DeclareMathSymbol{\FORALL}   {\mathord}{symbols}{"38}
\DeclareMathSymbol{\EXISTS}   {\mathord}{symbols}{"39}
\DeclareMathSymbol{\SUCHTHAT} {\mathbin}{symbols}{"01}
\DeclareMathAlphabet{\mathsc} {OT1}{cmr}{m}{sc}

\def\Forall#1#2{{\FORALL#1}: #2}
\def\Exists#1#2{{\EXISTS#1}: #2}
\DeclareSymbolFont{AMSb}{U}{msb}{m}{n}
\DeclareMathSymbol{\N}{\mathbin}{AMSb}{"4E}
\DeclareSymbolFontAlphabet{\mathbb}{AMSb}

\newcommand{\set}[1]{\ensuremath{\{#1\}}}

\newcommand{\SHOQ}{$\mathcal{SHOQ}$}

\newcommand{\roo}{\varepsilon}

\newcommand{\prule}[2]{\ensuremath{\mathit{#1}\gets\mathit{#2}}}
\newcommand{\nprule}[3]{\ensuremath{\mathit{#1}:\mathit{#2}\gets\mathit{#3}}}

\newenvironment{programn}{\[\begin{array}{rrll}}{\end{array}\]}
\newcommand{\nrule}[3]{\ensuremath{\mathit{#1}: & \mathit{#2} & \gets & \mathit{#3} \\}}

\newcommand{\naf}[1]{\ensuremath{\mathit{not}~ #1}}
\newcommand{\NAF}{\ensuremath{\mathit{not}}}

\newcommand{\HB}[1]{\ensuremath{\mathcal{B}_{#1}}}

\newcommand{\ground}[2]{{#1}_{#2}}

\newcommand{\preds}[1]{\ensuremath{\mathit{preds}(#1)}}

\newcommand{\upreds}[1]{\ensuremath{\mathit{upreds}(#1)}}
\newcommand{\bpreds}[1]{\ensuremath{\mathit{bpreds}(#1)}}
\newcommand{\vars}[1]{\ensuremath{\mathit{vars}(#1)}}

\newcommand{\nexptimenp}[1]{\ensuremath{\textsc{nexptime}^{\textsc{np}}}}

\newcommand{\nexptime}[1]{\ensuremath{\textsc{nexptime}}}

\newcommand{\CS}{\ensuremath{\mathit{CS}}}

\newenvironment{proofsketch}{{\it Proof Sketch. }}{}

\newcommand{\posi}[1]{\ensuremath{{#1}^{+}}}
\newcommand{\nega}[1]{\ensuremath{{#1}^{-}}}

\newcommand{\cts}[1]{\ensuremath{\mathit{cts}{(#1)}}}

\newcommand{\dlpluslog}{\ensuremath{\mathcal{DL}\text{+}\mathit{log}}}

\newcommand{\EF}{\ensuremath{\mathit{EF}}}

\newcommand{\ES}{\ensuremath{\mathit{ES}}}

\begin{document}

\title{An Optimization for Reasoning with Forest Logic Programs
\thanks{This work is partially supported by the Austrian Science Fund (FWF) under the projects P20305 and P20840, and by the European Commission under the project OntoRule (IST-2009-231875).}
}
\author{CRISTINA FEIER, STIJN HEYMANS}
\institute{Knowledge-Based Systems Group, Institute of Information Systems\\
Vienna University of Technology \\
Favoritenstrasse 9-11, A-1040 Vienna, Austria \\
\email{\{feier,heymans\}@kr.tuwien.ac.at}}

\maketitle

\begin{abstract}
Open Answer Set Programming (OASP) is an attractive framework for integrating ontologies and rules. In general OASP is undecidable. In previous work we provided a tableau-based algorithm for satisfiability checking w.r.t. forest logic programs, a decidable fragment of OASP, which has the forest model property. In this paper we introduce an optimized version of that algorithm achieved by means of a knowledge compilation technique. So-called unit completion structures, which are possible building blocks of a forest model, in the form of trees of depth 1, are computed in an initial step of the algorithm. Repeated computations are avoided by using these structures in a pattern-matching style when constructing a model. Furthermore we identify and discard redundant unit completion structures: a structure is redundant if there is another structure which can always replace the original structure in a forest model.
\end{abstract}

\section{Introduction}

Integrating Description Logics (DLs) with rules for the Semantic Web has received considerable attention with approaches such as \emph{Description Logic Programs} \cite{grosof}, \emph{DL-safe rules} \cite{motik}, \dlpluslog{} \cite{rosati-kr2006}, \emph{dl-programs} \cite{eiter-ai2008}, \emph{Description Logic Rules} \cite{Krotzsch+Rudolph+Hitzler-DLRules:2008}, and Open Answer Set Programming (OASP) \cite{heymans-acm2008}. OASP is a formalism which combines attractive features from the Logic Programming (LP) and the DL world. The syntax and semantics of OASP build upon the syntax and semantics of Answer Set Programming (ASP) \cite{gelfond88stable}: there is a rule-based syntax with a \emph{negation as failure} operator which is interpreted via a stable model semantics, but unlike the LP setting,
an open domain semantics, like it is common in the DL world, is employed. This allows for stating generic knowledge, without the need to mention actual constants.

Several decidable fragments of OASP were identified by syntactically restricting the shape of logic programs, while carefully safe-guarding enough expressiveness for integrating rule- and ontology-based knowledge. A notable fragment is that of \emph{Forest Logic Programs (FoLPs)} \cite{heymans-jal2007} that are able to simulate reasoning in the DL \SHOQ{}. 
FoLPs allow for the presence of only unary and binary predicates in rules which have a tree-like structure. A sound and complete algorithm for satisfiability checking of unary predicates w.r.t. FoLPs has been presented in \cite{feier+heymans-HybridReasoningForestLogicPrograms:09}. The algorithm exploits the forest model property of the fragment: if a unary predicate is satisfiable, than it is satisfied by a forest-shaped model, with the predicate checked to be satisfiable being in the label of the root of one of the trees composing the forest. It is essentially a tableau-based procedure which builds such a forest model in a top-down fashion.

In this paper we describe an optimization for reasoning with FoLPs in the form of a knowledge compilation technique. The technique consists in pre-computing all possible building blocks of the tableau, in the form of trees of depth 1, blocks which we call \emph{unit completion structures}. The original algorithm is used for computing the unit completion structures. The revised algorithm matches and appends such building blocks until a termination condition is met, like blocking or reaching a certain depth in the tableau expansion.
In general, not all unit completion structures have to be considered: inherent redundancy in a FoLP, like rules which are less general than others gives rise to redundancy among completion structures. A unit completion structure is redundant iff there is another simpler (less constrained) unit completion structure. The latter can replace the former in any forest model. We formalize this notion, making it possible to identify such redundant structures and discard them.

The paper is structured as follows: Section \ref{sec:prelim} contains preliminaries, like the OASP semantics and some notation, and Section \ref{sec:FLP} introduces the FoLP fragment. An overview of the original algorithm for reasoning with FoLPs is given in Section \ref{sec:algorithm}. The main results of the paper concerning the computation of non-redundant unit completion structures, and the revised algorithm, are presented in Section \ref{sec:opt}. Finally, Section \ref{sec:discussion} draws some conclusions and discusses future work.

\section{Preliminaries}
\label{sec:prelim}

We recall the open answer set semantics \cite{heymans-acm2008}. \emph{Constants} $a,b,c,\ldots$, \emph{variables} $X,Y,\ldots$, \emph{terms} $s,t,\ldots$, and \emph{atoms} $p(t_1,\ldots,t_n)$ are as usual.  
A \textit{literal} is an atom $L$ or a negated atom $\naf{L}$.  We allow for \emph{inequality literals} of the form $s\neq t$, where $s$ and $t$ are terms. A literal that is not an inequality literal will be called a \emph{regular literal}. For a set $S$ of literals or (possibly negated) predicates, $\posi{S} = \set{a \mid a \in S}$ and $\nega{S} = \set{a \mid \naf{a} \in S}$. For a set $S$ of atoms, $\naf{S} = \set{\naf{a} \mid a \in S}$.  For a set of (possibly negated) predicates $S$, $S(X) = \{a(X) \mid a \in S\}$ and $S(X,Y)=\{a(X,Y) \mid a \in S\}$. For a predicate $p$, $\pm p$ denotes $p$ or $\naf p $, whereby multiple occurrences of $\pm p$ in the same context will refer to the same symbol (either $p$ or $\naf p$).

A \textit{program} is a countable set of rules \prule{\alpha}{\beta}, where $\alpha$ is a finite set of regular literals and $\beta$ is a finite set
of literals. The set $\alpha$ is the \textit{head} and represents a disjunction, while $\beta$ is the \textit{body} and represents a conjunction.  If $\alpha =\emptyset$, the rule is called a \textit{constraint}. A special type of rules with empty bodies, are so-called \emph{free rules} which are rules of the form: ${q(t_1,\ldots,t_n)\lor\naf{q(t_1,\ldots,t_n)}\gets}{}$, for terms $t_1,\ldots,t_n$; these kind of rules enable a choice for the inclusion of atoms in the open answer sets. We call a predicate $q$ \emph{free} if there is a ${q(X_1,\ldots,X_n)\lor\naf{q(X_1,\ldots,X_n)}\gets}{}$, with variables $X_1,\ldots,X_n$. Atoms, literals, rules, and programs that do not contain  variables are \textit{ground}. For a rule or a program $R$, let $\cts{R}$ be the constants in $R$, $\vars{R}$ its variables, and $\preds{R}$ its predicates with $\upreds{R}$ the unary and $\bpreds{R}$ the binary predicates. For every non-free predicate $q$ and a program $P$, $P_q$ is the set of rules of $P$ that have $q$ as a head predicate.  A \emph{universe} $U$ for $P$ is a non-empty countable superset of the constants in $P$: $\cts{P} \subseteq U$. We call $\ground{P}{U}$ the ground program obtained from $P$ by applying all possible substitutions of variables by elements of $U$ to every rule in $P$. Let $\HB{P}$ ($\mathcal{L}_P$) be the set of regular atoms (literals) that can be formed from a ground program $P$.

An \textit{interpretation} $I$ of a ground $P$ is a subset of $\HB{P}$. We write $I\models p(t_1,\ldots,t_n)$ if $p(t_1,\ldots,t_n) \in I$ and $I \models \naf{p(t_1,\ldots,t_n)}$ if $I\not\models p(t_1,\ldots,t_n)$. Also, for ground terms $s,t$, we write
$I\models s\neq t$ if $s \neq t$. For a set of ground literals $L$, $I\models L$ if $I \models l$ for every $l \in L$.  A ground rule $r:\prule{\alpha}{\beta}$ is \textit{satisfied} w.r.t. $I$, denoted $I \models r$, if $I \models l$ for some $l \in \alpha$ whenever $I \models \beta$.  A ground constraint \prule{}{\beta} is satisfied w.r.t. $I$ if $I \not\models \beta$.

For a positive ground program $P$, i.e., a program without \NAF, an interpretation $I$ of $P$ is a \textit{model} of $P$ if $I$ satisfies every rule in $P$; it is an \textit{answer set} of $P$ if it is a subset minimal model of $P$. For ground programs $P$ containing \NAF, the \textit{GL-reduct} \cite{gelfond88stable} w.r.t. $I$ is defined as $P^I$, where $P^I$ contains \prule{\posi{\alpha}}{\posi{\beta}} for \prule{\alpha}{\beta} in $P$, $I \models \naf{\nega{\beta}}$ and $I \models \nega{\alpha}$.  $I$ is an \textit{answer set} of a ground $P$ if $I$ is an answer set of $P^I$.

A program is assumed to be a finite set of rules; infinite programs only appear as byproducts of grounding with an infinite universe. An \textit{open interpretation} of a program  $P$ is a pair $(U, M)$ where $U$ is a universe for $P$ and $M$ is an interpretation of $P_{U}$. An \textit{open answer set} of $P$ is an open interpretation $(U,M)$ of $P$ with $M$ an answer set of $\ground{P}{U}$. An $n$-ary predicate $p$ in $P$ is \emph{satisfiable} if there is an open answer set $(U,M)$ of $P$ s. t. $p(x_1,\ldots,x_n) \in M$, for some $x_1, \ldots, x_n \in U$.

We introduce notation for trees which extend those in \cite{Vardi98reasoningabout}. Let $\cdot$ be a concatenation operator between sequences of constants or natural numbers. A \emph{tree} $T$ with root $c$ ($T_c$), where $c$ is a specially designated constant, has as nodes sequences of the form $c \cdot s$, where $s$ is a (possibly empty) sequence of positive integers formed with the concatenation operator; for $x \cdot d\in T$, $d\in \N^{*}$, we have that $x\in T$. The set $A_{T}=\{(x,y) \mid x, y\in T, \Exists{n\in\N^{*}}{y=x \cdot n}\}$ is the set of arcs of a tree $T$. For $x,y\in T$, we say that $x <_T y$ iff $x$ is a prefix of $y$ and $x\neq y$.

A \emph{forest} $F$ is a set of trees $\{T_c \mid c \in C\}$, where $C$ is a set of distinguished constants. We denote with $N_F=\cup_{T\in F} T$ and $A_F=\cup_{T\in F}A_T$ the set of nodes and the set of arcs of a forest $F$, respectively. Let $<_F$ be a strict partial order relationship on the set of nodes $N_{F}$ of a forest $F$ where $x <_F y$ iff $x <_T y$ for some tree $T$ in $F$.  An extended forest $\EF$ is a tuple $(F, \ES)$ where $F =\{T_c \mid c\in C\}$ is a forest and $\mathit{ES}\subseteq N_F\times C$. We denote by $N_{\EF}=N_F$ the nodes of $\EF$ and by $A_{\EF}=A_F \cup \ES$ its arcs. So unlike a normal forest, an extended forest can have arcs from any of its nodes to any root of some tree in the forest.

Finally, for a directed graph $G$, $paths_G$ is the set of pairs of nodes for which there exists a path in $G$ from the first node in the pair to the second one.

\section{Forest Logic Programs}\label{sec:FLP}

\emph{Forest Logic Programs (FoLPs)} \cite{heymans-jal2007} are logic programs with tree-shaped rules which allow for constants and for which satisfiability checking under the open answer set semantics is decidable.

\begin{definition}\label{def:FoLP} A \emph{forest logic program (FoLP)} is a program with only unary and binary predicates, and such that a rule is either
a \emph{free rule}
\prule{a(s)\lor \naf{a(s)}}{} or \prule{f(s,t)\lor \naf{f(s,t)}}{},
where $s$ and $t$ are terms such that if $s$ and $t$ are both variables, they are different,
a \textit{unary rule}
\begin{equation}\label{eq:unary}
r:\prule{a(s)}{\beta(s), (\gamma_m(s,t_m),\delta_m(t_m))_{1\leq m \leq k},\psi}
\end{equation}
where $s$ and $t_m$, $1\leq m \leq k$, are terms (again, if both $s$ and $t_m$ are variables, they are different; similarly for $t_i$ and $t_j$), where
\begin{itemize}
\item $\psi\subseteq \bigcup_{1\leq i\neq j\leq k}\set{t_i\neq t_j}$ and $\set{\neq} \cap \gamma_m = \emptyset$ for $1\leq m \leq k$,
\item $\Forall{t_i\in \vars{r}}{{\gamma_i^+}\neq\emptyset}$, i.e., for variables $t_i$ there is a positive atom that connects $s$ and $t_i$,
\end{itemize} or a \emph{binary rule}
\begin{equation}\label{eq:binary}
\prule{f(s,t)}{\beta(s), \gamma(s,t), \delta(t)}
\end{equation}
with $\set{\neq}\cap \gamma = \emptyset$ and $\posi{\gamma} \neq \emptyset$ if $t$ is a variable ($s$ and $t$ are different if both are variables),
or a \textit{constraint}
\prule{}{a(s)} or\prule{}{f(s,t)}
where $s$ and $t$ are different if both are variables).
\end{definition}

The following program $P$ is a FoLP which says that an individual is a special member of an organization (\emph{smember}) if it has the support of another special member: \emph{rule} $r_1$, or if it has the support of two regular members of the organization (\emph{rmember}): \emph{rule} $r_2$. The binary predicate \emph{support} which describes the `has support' relationship is free. Nobody can be at the same time both a special member or a regular member: \emph{constraint} $r_4$. Two particular regular members are $a$ and $b$: \emph{facts} $r_5$ and $r_6$.

\begin{example}\label{ex:FoLP}
\small
\begin{programn}
\label{folp}
\nrule{r_1}{smember(X)}{support(X,Y), smember(Y)}
\nrule{r_2}{smember(X)}{support(X,Y), rmember(Y),}
&&&\ensuremath{support(X,Z), rmember(Z), Y \neq Z}\\
\nrule{r_3}{support(X, Y) \lor \naf{support(X,Y)}} {}
\nrule{r_4} {} {smember(X), rmember(X)}
\nrule{r_5}{rmember(a)}{}
\nrule{r_6}{rmember(b)}{}
\end{programn}
\normalsize
\end{example}

As their name suggests FoLPs have the \emph{forest model property}:

\begin{definition}\label{def:forest-sat1}
Let $P$ be a program.
A predicate $p\in \upreds{P}$ is \emph{forest satisfiable}\index{forest satisfiable} w.r.t. $P$ if there is an open answer set $(U,M)$ of $P$ and there is an extended forest
$\EF \equiv (\set{T_{\roo}} \cup \set{T_a \mid a \in \cts{P}},\ES)$, where $\roo$ is a constant, possibly one of the constants appearing in $P$
, and a labeling function
$\mathcal{L}:\set{T_{\roo}} \cup \set{T_a \mid a \in \cts{P}}\cup A_{\EF} \to 2^{\preds{P}}$ s. t.
\begin{itemize}
\item $U = N_{\EF}$, and 
\item $p \in \mathcal{L}(\roo)$,
\item $z \cdot i \in T\in \EF$, $i > 0$, iff there is some $f(z,z\cdot i) \in M$, $z\in T$, and
\item for $y \in T\in \EF$, $q \in \upreds{P}$, $f \in \bpreds{P}$,
we have that
	\begin{itemize}
	\item $q(y) \in M$ iff $q \in \mathcal{L}(y)$, and
	\item $f(y,u) \in M$ iff $(u = y\cdot i \lor u \in \cts{P}) \land f \in \mathcal{L}(y,u)$.
	\end{itemize}
\end{itemize}

We call such a $(U,M)$ a \emph{forest model} and a program $P$ has the \emph{forest model property} if the following property holds: if $p \in \upreds{P}$ is satisfiable w.r.t. $P$ then $p$ is forest satisfiable w.r.t. $P$.
\end{definition}

Consider the FoLP $P$ introduced in Example \ref{ex:FoLP}. The unary predicate $smember$ is forest satisfiable w.r.t. $P$: $(\{a,b,x\},$ $ \{rmember(a),$ $ rmember(b), $ $support (x,a),$ $support(x,b),$ $smember(x)\})$ is a forest model in which $smember$ appears in the label of the (anonymous) root of one of the trees in the forest (see Figure 1). Note that in the ordinary LP setting, where one restricts the universe to the Herbrand universe, $smember$ is not satisfiable.

\vspace{-8mm}
\small
\begin{center}
\label{fig:forestmodel}
\begin{figure}
\begin{tikzpicture}[level distance=3.5cm, >=stealth']
    \node(x){$x$}
          child[grow=left]{
                node (a) {$a$}
		        edge from parent [->]
       	         	node[below,near end, xshift=0.5cm] {$\{support\}$}
          }
	  child[grow=right]{
		node (b) {$b$}
		        edge from parent [->]
       	         	node[below,near end, xshift=-0.5cm] {$\{support\}$}
          };
    \node[right, xshift=-1.2cm, yshift=0.35cm] at (x.east) {$\{smember\}$};
    \node[right, xshift=-2.1cm] at (a.east) {$\{rmember\}$};
    \node[right, xshift=-0.2cm] at (b.east) {$\{rmember\}$};
    \end{tikzpicture}
    \vspace{-2mm}
    \caption{A Forest Model for $P$}
\end{figure}
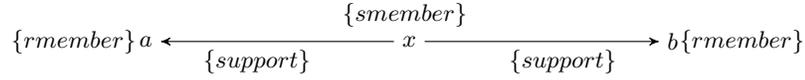
\end{center}
\normalsize
\vspace{-18mm}


\section{An Algorithm for Forest Logic Programs}\label{sec:algorithm}

In this section, we give an overview of the tableau algorithm for satisfiability checking for FoLPs introduced in \cite{feier+heymans-HybridReasoningForestLogicPrograms:09}. For technical details we refer the reader to the original paper. We use as a running example the FOLP from Example \ref{ex:FoLP}. Constraints are not treated explicitly in the algorithm as they can be simulated using unary rules. As such, the constraint $\nprule{r_4}{}{smember(X), rmember(X)}$ in Example \ref{ex:FoLP} is replaced with $\nprule{r'{_4}}{co(X)} {\naf{co}(X), smember(X), rmember(X)}$, with $\mathit{co}$ a new predicate.




The basic data structure used by the algorithm is a so-called  \emph{completion structure}. It contains an extended forest $EF$, whose set of nodes constitutes the universe of the model, and a labeling function $\ct$ (\textit{content}), which assigns to every node, resp. arc of $EF$, a set of possibly negated unary, resp. binary predicates. The presence of a predicate symbol $p/\naf p$ in the content of some node or arc $x$ indicates the presence/absence of the atom $p(x)$ in the open answer set. The presence (absence) of an atom in the open answer set is justified by imposing that the body of at least one ground rule which has the respective atom in the head is satisfied (no body of a rule which has the respective atom in the head is satisfied). In order to keep track which (possibly negated) predicate symbols in the content of some node or arc have already been expanded a completion structure contains also a so-called status function \st. Furthermore, in order to ensure that no atom in the partially constructed open answer set is circularly motivated, i.e. the atoms are well-supported \cite{fages-NewfixpointSem:91}, a graph $G$ which keeps track of dependencies between atoms in the (partial) model is maintained.

\begin{definition}\label{def:completionstructure} An $\mathcal{A}_1$-\textit{completion structure for a FoLP $P$}\footnote{We use the prefix $\mathcal{A}_1$ to denote completion structures computed using this original algorithm as opposed to completion structures computed using the optimised algorithm described in the next section for which we will use the prefix $\mathcal{A}_2$.} is a tuple $\langle \EF,$ $\ct,$ $\st,$ $G \rangle$ 
where:
\begin{itemize}
\item $\EF=\langle F,\ES \rangle$ is an extended forest,
\item $\ct:N_{\EF} \cup A_{\EF} \to 2^{preds (P) \cup \naf{(preds(P))}}$ is the `content' function,
\item $\st:\{(x, \pm q) \mid \pm q \in \ct(x), x \in N_{\EF} \cup A_{\EF}\} \to \{\expa, \unexpa\}$
is the `status' function,
\item $G=\grap{V}{A}$ is a directed graph which has as vertices atoms in the answer set in construction: $V \subseteq \HB{\ground{P}{N_{\EF}}}$.
\end{itemize}

An \emph{initial $\mathcal{A}_1$-completion structure} for checking satisfiability of a unary predicate $p$ w.r.t. a FoLP $P$ is a completion structure $\langle \EF,$ $\ct,$ $\st,$ $G \rangle$ with $\EF=(F,\emptyset)$, $F=\set{T_{\roo}} \cup \set{T_a \mid a \in \cts{P}}$, where $\roo$ is a constant, possibly in $\cts{P}$, $T_x=\{x\}$, for $x \in \{\roo\} \cup \cts{P}$, $G=\grap{V}{\emptyset}$, $V=\{p(\roo)\}$, and $\ct(\roo)=\{p\}$, $\st(\roo,p)=\unexpa$.
\end{definition}

An extended forest is initialized with single-node trees having as roots constants from $P$ and, possibly, a new single-node tree with anonymous root. 
The forest model from Figure 1 has been evolved from an initial completion structure which has as $\roo$, the root element where $smember$ has to be satisfied, the anonymous individual, $x$. There are two other single-node trees: $T_a$ and $T_b$. The predicate $smember$ in the content of $x$ is marked as unexpanded and $G$ is a graph with a single vertex $smember(x)$.

\vspace{1mm}
\small
\begin{center}
\begin{tabular}{ll}
$EF$:&
\begin{tikzpicture}[level distance=3.5cm, >=stealth']
    \node[yshift=-0.35cm](x){$x$};
    \node[xshift=-3.5cm](a) at (x.west) {$a$};
    \node[xshift=3.5cm](b) at (x.east) {$b$};
    \node[right, xshift=-1.2cm, yshift=0.35cm] at (x.east) {$\{smember^u\}$};
    \node[right, xshift=-0.2cm] at (a.east) {$\{\}$};
    \node[right, xshift=-0.2cm] at (b.east) {$\{\}$};
    \end{tikzpicture}
     \vspace{2mm} \\ $V$: & $smember(x)$ \\
$A$: & $\emptyset$
\end{tabular}
\end{center}
\normalsize

An initial $\mathcal{A}_1$-completion structure for checking the satisfiability of a unary predicate $p$ w.r.t. a FoLP $P$ is evolved by means of \emph{expansion rules} to a complete clash-free structure that corresponds to a finite representation of an open answer set in case $p$ is satisfiable w.r.t. $P$. \emph{Applicability rules} govern the application of the expansion rules.

\subsection{Expansion Rules}\label{sec:expansionrules}


In the following, for a completion structure $\langle \EF,$ $\st,$ $\ct,$  $G \rangle$, let $x \in N_{\EF}$ and $(x,y)\in A_{\EF}$ be the node, resp. arc, under consideration.

\textbf{\emph{(i) Expand unary positive}}. For a unary positive (non-free) $p \in \ct(x)$ s. t. $\st(x,p)=\unexpa$,
choose a unary rule $r\in P_p$
 for which $s$, the head term, matches $x$; ground this rule by substituting $s$ with $x$, in case $s$ is a variable, and every successor term  $t_m$ which is a variable with a successor of $x$ in $EF$ s. t. the inequalities in $\psi$ are satisfied (if needed one can introduce new successors of $x$ in $EF$, either as successors of $x$ in $T$, where $x \in T$, or in the form of constants from $P$). We motivate the presence of $p(x)$ in the open answer set by enforcing its body to be satisfied by inserting appropriate (possibly negated) predicate symbols in the contents of nodes/arcs of the structure. The newly inserted predicate symbols are marked as unexpanded and $G$ is updated, by adding arcs from $p(x)$ to every body atom.


In our example, $smember$ is unexpanded in the initial completion structure. Rule $r_2$ is chosen to motivate the presence of $smember(x)$ in the open answer set. It is grounded by substituting $X$ with $x$, and $Y_1$ and $Y_2$ with $a$ and $b$, respectively:  $\prule{smember(x)}{support(x,a)}$ $, rmember(x,a),$ $support(x,b), rmember(x,b)$. We enforce the body of this ground rule to be true and obtain the following completion structure (note also that $G$ has been updated):

\small
\begin{center}
\begin{tabular}{ll}
$EF$:\hspace{1mm}& \hspace{1mm}
\begin{tikzpicture}[level distance=3.5cm, >=stealth']
    \node(x)[yshift=-0.35cm]{$x$}
          child[grow=left]{
                node (a) {$a$}
		        edge from parent [->]
       	         	node[below,near end, xshift=0.5cm] {$\{support^{u}\}$}
          }
	  child[grow=right]{
		node (b) {$b$}
		        edge from parent [->]
       	         	node[below,near end, xshift=-0.5cm] {$\{support^{u}\}$}
          };
    \node[right, xshift=-1.2cm, yshift=0.35cm] at (x.east) {$\{smember^{e}\}$};
    \node[right, xshift=-2.2cm] at (a.east) {$\{rmember^{u}\}$};
    \node[right, xshift=-0.2cm] at (b.east) {$\{rmember^{u}\}$};
    \end{tikzpicture}
     \vspace{2mm}
 \\
    $V:$\hspace{1mm} &\hspace{1mm}$smember(x), support(x,a), support(x, b), rmember(x, a), rmember(x,b)$
\\
    $A:$\hspace{1mm} &\hspace{1mm}$smember(x) \rightarrow support(x,a), smember(x) \rightarrow support(x,b),$\\
    &\hspace{1mm}$smember(x) \rightarrow rmember(x,a),$ $smember(x) \rightarrow rmember(x,b)$
\end{tabular}
    \end{center}
\vspace{-1mm}
\normalsize

 All currently unexpanded predicates, i.e., $support$ in the content of arcs $(x,a)$ and $(x,b)$, and $rmember$ in the content of nodes $a$ and $b$,  can be trivially expanded as $support$ is a free predicate and $r_5$ and $r_6$ are facts. However one still has to ensure that the structure constructed so far can be extended to an actual open answer set, i.e., it is consistent with the rest of the program. The following expansion rule takes care of this.

\textbf{\emph{(ii) Choose a unary predicate}}. If all predicates in $\ct(x)$ and in the contents of x's outgoing edges are expanded and there are still unary predicates $p$ which do not appear in $\ct(x)$, pick such a $p$ and inject either $p$ or $\naf p$ in $\ct(x)$. The intuition is that one has to explore all unary/binary predicates at every node/arc as some predicate which is not reachable by dependency-directed expansion can prohibit the extension of the partially constructed model to a full model. Consider the simple case where there is a predicate $p$ defined only by the rule: $\prule{p}{\naf p}$ and $\pm p$ does not appear in the body of any other rule. The program is obviously inconsistent, but this cannot be detected without trying to prove that $p$ is or is not in the open answer set.

In our example, one does not know whether $co$ or $\naf{co}$ belongs to $\ct(x)$. We choose to inject $\naf{co}$ in $\ct(x)$ and mark it as unexpanded.


\textbf{\emph{(iii) Expand unary negative}}. Justifying a negative unary predicate $\naf p \in \ct(x)$ means refuting the body of every ground rule which defines $p(x)$, or in other words refuting at least a literal from the body of every ground rule which defines $p(x)$.
For more technical details concerning this rule we refer the reader to \cite{feier+heymans-HybridReasoningForestLogicPrograms:09}.

In our example, the unexpanded predicate in $\ct(x)$, $\naf{co}$, is defined by one rule, $r'{_4}$, whose only possible grounding is $\prule{co(x)} {\naf{co}(x), smember(x), rmember(x)}$. Refuting the body of this rule amounts to inserting $\naf{rmember}$ in $\ct(x)$ ($smember$ and $\naf{co}$ are already part of the content of that node). At its turn, the presence of $\naf{rmember}$ in $\ct(x)$ has to be motivated by using the expand unary negative rule, and the process goes on. Finally, we obtain a completion structure in which no expansion rule is further applicable and which represents exactly the forest model from Figure 1 ($smember$ and $rmember$ are abbreviated with $sm$ and $rm$, respectively):

\vspace{-3mm}
\small
\begin{center}
\begin{tabular}{ll}
$EF$:& \begin{tikzpicture}[level distance=4.2cm, >=stealth']
    \node(x)[yshift=-0.35cm, xshift=-2.0cm]{$x$}
          child[grow=left]{
                node (a) {$a$}
		        edge from parent [->]
       	         	node[below,near end, xshift=1.7cm] {$\{support\}$}
          }
	  child[grow=right]{
		node (b) {$b$}
		        edge from parent [->]
       	         	node[below,near end, xshift=-1.7cm] {$\{support\}$}
          };
    \node[right, xshift=-1.7cm, yshift=0.35cm] at (x.east) {$\{sm, \naf{rm}, \naf{co}\}$};
    \node[right, xshift=-1.7cm, yshift=-0.35cm] at (a.east) {$\{rm, \naf{sm}, \naf{co}\}$};
    \node[right, xshift=-1.7cm, yshift=-0.35cm] at (b.east) {$\{rm, \naf{sm}, \naf{co}\}$};
    \end{tikzpicture}
     \vspace{2mm}
 \\
    $V:$&$sm(x), support(x,a), support(x, b), rm(x, a),$ $rm(x,b)$
\\
    $A:$&$sm(x) \rightarrow support(x,a), sm(x) \rightarrow support(x,b),$
    $sm(x) \rightarrow rm(x,a),$ $sm(x)$ $\rightarrow rm(x,b)$
\end{tabular}
    \end{center}
\vspace{0mm}
\normalsize

Similarly to rules (i), (ii), and (iii) we define the expansion rules for binary predicates: (iv) \emph{Expand binary positive}, (v) \emph{Expand binary negative}, and (vi) \emph{Choose binary}.

\subsection{Applicability Rules}\label{sec:appl}

The applicability rules restrict the use of the expansion rules.

\textbf{\emph{(vii) Saturation}.} A node $x \in N_{\EF}$ is \emph{saturated} if for all ${p \in \upreds{P}}$, $p \in \ct(x)$ or $\naf p \in \ct(x)$, and no $\pm q \in \ct(x)$ can be expanded with rules (i-iii), and for all ${(x,y) \in A_{\EF}}$ and $p \in \bpreds{P}$, $p \in \ct(x,y)$ or $\naf p \in \ct(x,y)$, and no $\pm f \in \ct(x,y)$ can be expanded with (iv-vi). No expansions should be performed on a node from $N_{\EF}$ which does not belong to $\cts{P}$ until its predecessor is saturated.

\textbf{\emph{(viii) Blocking}.} A node $x \in N_\EF$ is \emph{blocked} if there is an ancestor $y$ of $x$ in $F$, $y<_F x$, $y \not \in \cts{P}$, s. t. $\ct(x) \subseteq \ct(y)$ and the set $paths_G(y,x)=\{(p,q) \mid (p(y), q(x)) \in paths_G \wedge q \mbox{ is not free}\}$ is empty. We call $(y,x)$ a \emph{blocking pair}. No expansions can be performed on a blocked node. One can notice that subset blocking is not enough for pruning the tableau expansion. Every atom in the open answer set has to be finitely motivated \cite[Theorem 2]{heymans-amai2006}: in order to ensure that, one has to check that there is no dependency in $G$ between an atom formed with the blocking node and an atom formed with the blocked node. The extra condition makes the blocking rule insufficient to ensure the termination of the algorithm. The following applicability rule ensures termination.

\begin{example}\label{ex:blocking}
Consider a restricted version of $P$ from Example \ref{ex:FoLP} which contains only rules $r_1$, and $r_3$. By checking satisfiability of $smember$ w.r.t. the new program one obtains the following completion structure:\newline
\small
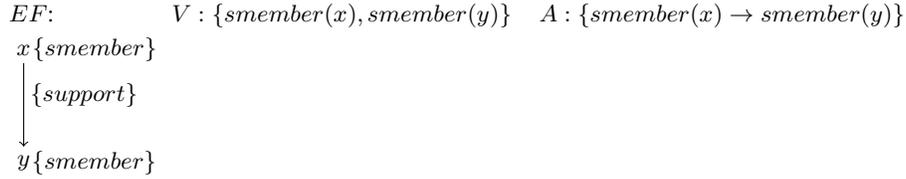
\begin{center}
\begin{tabular}{lll}
$EF$: \hspace{10mm}
&
$V:\{smember(x), smember(y)\}$&\hspace{2mm}
$A:\{smember(x) \to smember(y)\}$
\\
\begin{tikzpicture}[ auto]
    \node (x) {$x$}
          child{
                node (y) {$y$}
		edge from parent [->]
		};
    	\node[yshift=-0.6cm, xshift=0.8cm] {$\{support\}$};
     \node[right, xshift=-0.2cm] at (x.east) {$\{smember\}$};
    \node[right, xshift=-0.2cm] at (y.east) {$\{smember\}$};
\end{tikzpicture}
&&
\end{tabular}
\end{center}
\normalsize
While the contents of nodes $x$ and $y$ are identical, they do not form a blocking pair as there is an arc in $G$ between $smember(x)$ and $smember(y)$: unfolding the structure (justifying $y$ similarly as $x$) would lead to an infinite chain: $smember(x), $ $smember(y), $ $smember(z), \ldots,$ in the atom dependency graph of the grounded program.
\end{example}


\textbf{\emph{(ix) Redundancy}.} A node $x \in N_\EF$ is \emph{redundant} if it is saturated, it is not blocked, and there are $k$ ancestors of $x$ in $F$, $(y_i)_{1 \leq i \leq k}$, with $k=2^p(2^{p^2}-1)+3$, and $p=|\upreds{P}|$, s. t. $\ct(x)=\ct(y_i)$. In other words, a node is redundant if it is not blocked and it has $k$ ancestors with content equal to its content: any forest model of a FoLP $P$ which satisfies $p$ can be reduced to another forest model which satisfies $p$ and has at most $k+1$ nodes with equal content on any branch of a tree from the forest model, and furthermore the $(k+1)st$ node, in case it exists, is blocked \cite{feier+heymans-HybridReasoningForestLogicPrograms:09}. One can thus search for forest models only of the latter type. As such the detection of a redundant node constitutes a clash and stops the expansion process.


\subsection{Termination, Soundness, Completeness, Complexity Results}

An $\mathcal{A}_1$-completion structure is \emph{contradictory} if for some $x \in N_ {\EF}/A_{\EF}$ and $p \in \upreds{P}/$ $\bpreds{P}$, $\{p, \naf p\} \subseteq \ct(x)$. An $\mathcal{A}_1$-completion structure for a FoLP $P$ and a $p\in\upreds{P}$ is \emph{complete} if it is a result of applying the expansion rules to the initial completion structure for $p$ and $P$, taking into account the applicability rules, s. t. no expansion rules can be further applied.

Also, a complete $\mathcal{A}_1$-completion structure  $\CS = \langle \EF,$ $\ct,$ $\st,$ $G \rangle$ is $\mathcal{A}_1$-\emph{clash-free} if:
 (1) \CS{} is not contradictory
 (2) $\EF$ does not contain redundant nodes
 (3) $G$ does not contain cycles (4) there is no $p \in \upreds{P}/\bpreds{P}$ and $x \in N_ {\EF}/A_{\EF}$, $x$ unblocked, s.t. $p \in \ct(x)$, and $\st(x,p)=\unexpa$.

It has been shown that an initial $\mathcal{A}_1$-completion structure for a unary predicate $p$ and a FoLP $P$ can always be expanded to a complete $\mathcal{A}_1$-completion structure (\emph{termination}), that, if $p$ is satisfiable w.r.t. $P$, there is a complete clash-free $\mathcal{A}_1$-completion structure (\emph{soundness}), and, finally, that, if there is a complete clash-free $\mathcal{A}_1$-completion structure, $p$ is satisfiable w.r.t. $P$ (\emph{completeness}).

In the worst case the algorithm runs in double exponential time, and a complete completion structure has a double exponential number of nodes in the size of the program. The high complexity is mostly due to the fact that blocking is not enough to ensure termination, and that, in particular, ``anywhere blocking''\cite{Motik+Shearer+Horrocks:DLReasHypertableaux} cannot be used as a termination technique. As already explained this peculiarity appears as a result of adopting a minimal model semantics.

\section{Optimized Reasoning with FoLPs}
\label{sec:opt}

This section presents a knowledge compilation technique for reasoning with FoLPs together with an algorithm which makes use of this pre-compiled knowledge. The main idea is to capture all possible local computations, which are typically performed over and over again in the process of saturating the content of a node, by pre-computing all possible completion structures of depth 1 using the original algorithm described in the previous section.
In the new algorithm, saturating the content of a node reduces to picking up one of the pre-computed structures which satisfies the existing constraints regarding the content of that node and appending the structure to the completion in construction: such constraints are sets of unexpanded (possibly negated) predicates which are needed to motivate the presence/absence in the open answer set of atoms constructed with the current node and the node above it.

Picking up a certain unit completion structure to saturate a node can impose strictly more constraints on the resulted structure than picking another unit completion structure with the same root content. Such constraints refer to: (1) the contents of the successor (non-blocked) nodes in a unit completion structure; (2)  the paths from an atom formed with the root node of a unit completion to an atom formed with a successor node of such a completion -- the more paths there are the harder blocking becomes. We discard such structures which are strictly more constraining than others, as they can be seen as redundant building blocks for a model.

The rest of the section formalizes and exemplifies these notions.

\subsection{Unit Completion Structures}

As mentioned in the introduction of this section, the intention is to obtain all completion structures of depth 1 which can be used as building blocks in our algorithm. We call such structures \emph{unit completion structures}. The skeleton of such a structure, is a so-called \emph{initial unit completion structure}. If they are to be used as building blocks in the algorithms, unit completion structures have to have as backbones trees of depth 1, and not forests. Hence, an initial unit completion structure is defined as a tree (unlike its counterpart notion from the previous section, initial completion structure, which is defined as a forest) with a single node, the root, which is either an anonymous constant or one of the constants already present in the program. The content of the root is empty.

\begin{definition}\label{def:initunitcomplstr}
An \emph{initial unit completion structure} for a FoLP $P$ is a completion structure $\langle \EF,$ $\ct,$ $\st,$ $G\rangle$ with $\EF=(F,\ES)$, $F=\set{T_{\roo}}$, where $\roo$ is a constant, possibly in $\cts{P}$, $T_{\roo}=\{\roo\}$, $\ES=\emptyset$, $G=\grap{V}{A}$, $V=\emptyset$, $A=\emptyset$, and $\ct(\roo)=\emptyset$.
\end{definition}

A unit completion structure captures a possible local computation: that is, it is obtained as an expansion of an initial unit completion structure, to a tree of depth 1.

\begin{definition}\label{def:unitcomplstr}
A \emph{unit completion structure} $\langle \EF,$ $\ct,$ $\st,$ $G \rangle$ for a FoLP $P$, with $\EF=(\set{T_{\roo}},\ES)$,  is an $\mathcal{A}_1$-completion structure derived from an initial unit completion structure by application of the expansion rules (i)-(vi) described in Section \ref{sec:expansionrules}, according to the applicability rules introduced in Section \ref{sec:appl}, which has been expanded such that $\roo$ is saturated and for all $s$ such that $\roo \cdot s \in T_{\roo}$, and for all $\pm p \in \ct(\roo \cdot s)$, $\st(\pm p, \roo \cdot s)=\unexpa$.\footnote{The status function is relevant only in the definition/construction of a unit completion structure, but not in the context of using such structures. As such, we will denote a unit completion structure in the following as a triple $\langle \EF,$ $\ct,$ $G \rangle$.}
\end{definition}

\begin{example}\label{ex:threeucps}

Consider the program $Pr$:

\begin{programn}
\label{folp2}
\nrule{r_1}{p(X)}{\naf{p(X)}}
\nrule{r_2}{p(X)}{f(X,Y), \naf{q(Y)}}
\nrule{r_3}{p(X)}{f(X,Y), p(Y)}
\nrule{r_4}{p(X)}{f(X,Y), \naf{q(Y)}, p(Y)}
\nrule{r_5}{q(X)}{f(X,Y), \naf{p(Y)}}
\nrule{r_6}{f(X,Y) \lor \naf{f(X,Y)}}{}
\end{programn}

\end{example}
Figure 2 depicts three unit completion structures for $Pr$. They all have the same content for the root node: $\{p, \naf{q}\}$. The presence of $p$ in the content of the root node has been motivated in the first structure by means of rule $r_4$, in the second structure by means of rule $r_3$, and in the third structure by means of rule $r_2$. The different ways to derive $p$ lead to different sets of arcs in the dependency graphs belonging to each structure. On the other hand, to motivate that $\naf{q}$ is in the content of the root node, in each case it was shown that the body of $r_5$ grounded such that $X$ is instantiated as the root node and $Y$ as the successor node is not satisfied, or more concretely the presence of $p$ in the content of the successor node was enforced in each case ($\naf{f}$ could not be used to invalidate the triggering of the rule as $f$ was already present in the content of the arc from the root node to the successor node in each case).

\begin{figure}
\vspace{3mm}
\small
\begin{center}
\begin{tabular}{l|l|l}
\hspace{1mm}
$UC_1:$&
\hspace{1mm}
$UC_2:$&
\hspace{1mm}
$UC_3:$
\\
\hspace{1mm}
\begin{tikzpicture}[ auto]
    \node (x) {$a$}
          child{
            node (y) {$a1$}
		    edge from parent [->]
            node[above,near end,xshift=0.3cm] {$\{f\}$}
            };
    \node[right, xshift=-0.2cm] at (x.east) {$\{p, \naf{q}\}$};
    \node[right, xshift=-0.2cm] at (y.east) {$\{p, \naf{q}\}$};
    \end{tikzpicture}
&
\hspace{1mm}
 \begin{tikzpicture}[ auto]
    \node (x) {$b$}
          child{
            node (y) {$b1$}
		    edge from parent [->]
            node[above,near end,xshift=0.3cm] {$\{f\}$}
            };
    \node[right, xshift=-0.2cm] at (x.east) {$\{p, \naf{q}\}$};
    \node[right, xshift=-0.2cm] at (y.east) {$\{p \}$};
    \end{tikzpicture}
&
\hspace{1mm}
\begin{tikzpicture}[ auto]
    \node (x) {$c$}
          child{
            node (y) {$c1$}
		    edge from parent [->]
            node[above,near end,xshift=0.3cm] {$\{f\}$}
            };
    \node[right, xshift=-0.2cm] at (x.east) {$\{p, \naf{q}\}$};
    \node[right, xshift=-0.2cm] at (y.east) {$\{p, \naf{q}\}$};
\end{tikzpicture}
\\
\hline
\hspace{2mm}
$G_1=(V_1, A_1)$&
\hspace{2mm}
$G_2=(V_2, A_2)$&
\hspace{2mm}
$G_3=(V_3, A_3)$
\\
& & \\

\hspace{2mm}
$V_1:p(a), p(a1), f(a, a1)$&
\hspace{2mm}
$V_2:p(b), p(b1), f(d, d1)$&
\hspace{2mm}
$V_3:p(c), p(c1), f(c, c1)$\\
&&\\
\hspace{2mm}
\multirow{2}{*}{$A_1:$}
$p(a) \rightarrow f(a,a1),$&
\hspace{2mm}
\multirow{2}{*}{$A_2:$}
$p(b)\rightarrow f(b,b1),$&
\hspace{2mm}
\multirow{2}{*}{$A_3:p(c)\rightarrow f(c,c1)$}\\
\hspace{8.5mm} $p(a) \rightarrow p(a1)$ &
\hspace{8.5mm} $p(b)\rightarrow p(b1)$\\
\end{tabular}
\vspace{2mm}
\label{figure:threeucps}
\caption{Three unit completion structures for $Pr$: $UC_1$, $UC_2$, and $UC_3$.}
\end{center}
\end{figure}
\normalsize
\vspace{-10mm}

One can notice that while the content of the successor node is included in the content of the root node in each of the cases, only for $UC_3$, the two nodes form a blocking pair as $paths_{G_3}(c, c1)=\emptyset$.

\begin{definition}\label{def:finalunitcomplstr}
A unit completion structure is \emph{final} iff all its successor nodes are blocked, or they have empty contents.
\end{definition}

\begin{proposition}\label{def:finalcompleteeq}
A final unit completion structure is a complete clash-free $\mathcal{A}_1$-completion structure.
\end{proposition}

In our example $UC_3$ is a final unit completion structure, and thus also a complete clash-free $\mathcal{A}_1$-completion structure.

\begin{proposition}\label{def:complexcomucp}
There is a deterministic procedure which computes all unit completion structures for a FoLP $P$ in the worst-case scenario in exponential time in the size of $P$.
\end{proposition}

\begin{proofsketch}
We consider the transformation of the non-deterministic algorithm described in Definition \ref{def:unitcomplstr} into a deterministic procedure. There are at most $2^n$ different values for the content of a saturated node, where $n=|upreds(P)|$. Justifying the presence of a predicate symbol $p$ in the content of a node takes in the worst case polynomial time, but there is an exponential number of choices to do this (an exponential number of possible groundings for every rule). Justifying the presence of a negated predicate symbol $\naf{p}$ in the content of a node takes in the worst case exponential time (all possible groundings of every rule $r \in P_p$ have to be considered), while at every step of the computation there is a polynomial number of choices. Overall, such a deterministic procedure runs in exponential time in the worst case scenario.
\qed
\end{proofsketch}

\subsection{Redundant Unit Completion Structures}

As seen in Example \ref{ex:threeucps}, there are unit completion structures with roots with equal content, but possibly different topologies, contents of the successor nodes and/or possibly different dependency graphs. As discussed in the introduction to this section it is worthwhile to identify structures which are strictly more constraining than others, in the sense that they impose more constraints on the content of the successor nodes of the structure and introduce more paths in the dependency graph as they can be discarded.
The following definition singles out such redundant structures.

\begin{definition}
A \emph{unit completion structure} $UC_1=\langle \EF_1,$ $\ct_1,$ $G_1\rangle$, with $\EF_1=(\{T_{\roo_1}\}, ES_1)$, is said to be \emph{redundant} iff there is another unit completion structure $UC_2=\langle \EF_2,$ $\ct_2,$ $G_2 \rangle$, with $\EF_2=(\{T_{\roo_2}\}, ES_2)$  s. t.:
\begin{itemize}

\item if $\roo_2 \in \cts{P}$, then $\roo_2=\roo_1$;

\item $\ct(\roo_1)=\ct(\roo_2)$;

\item if $\roo_2 \cdot s_1, \ldots, \roo_2 \cdot s_l$ are the non-blocked successors of $\roo_2$, there exist $l$ distinct successors $\roo_1 \cdot t_1, \ldots, \roo_1 \cdot t_l$ of $\roo_1$ such that:
\begin{itemize}
\item $\ct(\roo_2 \cdot s_i) \subseteq \ct(\roo_1 \cdot t_i)$, for every $1 \leq i \leq l$, and
\item $paths_{G_2}(\roo_2, \roo_2 \cdot s_i) \subseteq paths_{G_1}(\roo_1, \roo_1 \cdot t_i)$, for every $1 \leq i \leq l$,
\end{itemize}
with at least one inclusion being strict.
\end{itemize}
\end{definition}

Considering the previous example, one can see that $UC_1$, and $UC_2$ are redundant structures, while $UC_3$ is not, as $UC_1$ is more constraining than $UC_2$, and $UC_2$ at its turn is more constraining than $UC_3$.

\begin{proposition}\label{prop:redundantcomplex}
Computing the set of non-redundant unit completion structures for a FoLP $P$ can be performed in the worst case in exponential time in the size of $P$.
\end{proposition}

\begin{proofsketch}
The result follows from the fact that there is an exponential number of unit completion structures for a FoLP $P$ in the worst case scenario.\qed
\end{proofsketch}

\subsection{Reasoning with FoLPs Using Unit Completion Structures}

We define a new algorithm which uses the set of pre-computed non-redundant completion structures. We call this algorithm $\mathcal{A}_2$. As in the case of the previous algorithm, $\mathcal{A}_2$ starts with an initial $\mathcal{A}_2$-completion structure for checking satisfiability of a unary predicate $p$ w.r.t. a FoLP $P$ and expands this to a so-called $\mathcal{A}_2$-completion structure.

An \emph{$\mathcal{A}_2$-completion structure} $\langle \EF,$ $\ct,$ $\st,$ $G \rangle$ is defined similarly as an $\mathcal{A}_1$-completion structure, but the \emph{status} function has a different domain, the set of nodes of the forest: $\st:N_{EF} \to \{\expa, \unexpa\}$.

An \emph{initial $\mathcal{A}_2$-completion structure for a unary predicate $p$ and FoLP $P$} is defined similarly as an initial $\mathcal{A}_1$-completion structure for $p$ and $P$, but in this case every node in the extended forest is marked as unexpanded: $\st(x)=\unexpa$, for every $x \in N_{EF}$.

The difference in the definition of an  $\mathcal{A}_2$-completion structure compared to its $\mathcal{A}_1$ homonym is due to the fact that in this scenario nodes are expanded by matching their content with existent unit completion structures, and not predicates in the content of nodes. We make explicit the notion of matching the content of a node with a unit completion structure by introducing a notion of \emph{local satisfiability}:

\begin{definition}\label{def:unitcomplstrsatisf}
A unit completion structure $UC$ for a FoLP $P$,  $\langle \EF,$ $\ct,$ $G \rangle$, with $\EF=(\set{T_{\roo}},\ES)$, \emph{locally satisfies} a (possibly negated) unary predicate $p$ iff $p \in \ct(\roo)$. Similarly, $UC$ locally satisfies a set $S$ of (possibly) negated unary predicates iff $S \subseteq \ct(\roo)$.
\end{definition}

All three unit completions in Figure 2 locally satisfy the set $\{a, \naf{b}\}$. It is easy to observe that if a unary predicate $p$ is not locally satisfied by any unit completion structure $UC$ for a FoLP $P$ (or equivalently $\naf p$ is locally satisfied by every unit completion structure), $p$ is unsatisfiable w.r.t. $P$. However, local satisfiability of a unary predicate $p$ in every unit completion structure for a FoLP $P$ does not guarantee 'global' satisfiability of $p$ w.r.t. $P$ (as in the case of the program in Example \ref{ex:blocking} whose only unit completion structure was the one depicted in that example).

When building an $\mathcal{A}_2$-completion structure $CS=\langle \EF,$ $\ct,$ $\st,$ $G\rangle$, with $G=(V, A)$, for a FoLP $P$ by using unit completion structures as building blocks an operation commonly appears: the expansion of a node $x \in N_{\EF}$ by addition of a unit completion structure $UC=\langle \EF^{'},$ $\ct^{'},$ $G{'} \rangle$, with $\EF^{'}=(\{T_{\roo}\}, ES^{'})$ and $G^{'}=(V^{'}, A^{'})$, which locally satisfies $\ct(x)$, at $x$, given that its root matches with $x$ \footnote{An anonymous individual matches with any term, while a constant matches only with itself; thus, unit completion structures with roots constants can only be used as initial building blocks for trees with roots the corresponding constants.}. We call this operation $expand_{CS}(x, UC)$. Formally, its application updates $CS$ as follows:

\begin{itemize}
\item \st(x)=\expa,
\item $N_{\EF}=N_{\EF} \cup \{x \cdot s \mid \roo \cdot s \in T_{\roo}\}$,
\item $A_{\EF}=A_{\EF} \cup \{(x, x \cdot s)\mid (\roo, \roo \cdot s) \in A_{\EF^{'}}\}$,
\item $\ct(x)=\ct(\roo)$. For all $s$ such that $\roo \cdot s \in T_{\roo}$, $\ct(x \cdot s)= \ct(\roo \cdot s)$,
\item $V=V \cup \{p(x) \mid p \in \ct(\roo)\} \cup \{p(x \cdot s) \mid p \in \ct(\roo \cdot s)\}$,
\item $A=A \cup \{(p(\overline{z}), q(\overline{y})) \mid (p(z), q(y)) \in A^{'}\}$, where $\overline{\roo}=x$, and $\overline{\roo \cdot s}=x \cdot s$.

\end{itemize}

The revised algorithm has a new rule which we call \emph{Match}. This rule replaces the expansion rules (i)-(vi) and the applicability rule (vii) from the original algorithm.

\textbf{\emph{Match}}. For a node $x \in N_{EF}$: if $\st(x)=\unexpa$ choose a non-redundant unit completion structure $UC$  with root matching $x$ which satisfies $\ct(x)$ and perform $expand_{CS}(x, UC)$.

However, rules \emph{(viii) Blocking} and \emph{(ix) Redundancy}  are still used.

\begin{definition}\label{def:complcomplucp}
A \emph{complete $\mathcal{A}_2$-completion structure} for a FoLP $P$ and a $p\in\upreds{P}$, is an $\mathcal{A}_2$-completion structure that results from applying the rule \emph{Match} to an initial $\mathcal{A}_2$-completion structure for $p$ and $P$, taking into account the applicability rules (viii) and (ix), s. t. no other rules can be further applied.
\end{definition}

The local clash conditions regarding contradictory structures or structures which have cycles in the dependency graph $G$ are no longer relevant:

\begin{definition}\label{def:clashfreecomplucp}
A complete $\mathcal{A}_2$-completion structure  $\CS = \langle \EF,$ $\ct,$ $\st,$ $G \rangle$ is \emph{clash-free} if (1) $\EF$ does not contain redundant nodes (2) there is no node $x \in N_{EF}$, $x$ unblocked, s.t. $st(x)=\unexpa$.
\end{definition}

Termination follows from the usage of the blocking and of the redundancy rule:

\begin{proposition}\label{terminationucp}
An initial $\mathcal{A}_2$-completion structure for a unary predicate $p$ and a FoLP $P$ can always be expanded to a complete $\mathcal{A}_2$-completion structure. \end{proposition}

The algorithm is sound and complete:

\begin{proposition}\label{prop:soundnesscomplucp}
A unary predicate $p$  is satisfiable w.r.t. a FoLP $P$ iff there is a complete clash-free $\mathcal{A}_2$-completion structure.
\end{proposition}

\begin{proofsketch}
Soundness of $\mathcal{A}_2$ follows from soundness of $\mathcal{A}_1$: any completion structure computed using $\mathcal{A}_2$ could have actually been computed using $\mathcal{A}_1$ by replacing every usage of the \emph{Match} rule with the corresponding rule application sequence used by $\mathcal{A}_1$ to derive the unit completion structure which is currently appended to the structure.

Completeness of $\mathcal{A}_2$ follows from completeness of $\mathcal{A}_1$: any clash-free complete $\mathcal{A}_1$-completion structure can be seen as a complete clash-free $\mathcal{A}_2$-completion structure. It is essential here that the discarded unit completion structures were strictly more constraining than some other (preserved) unit completion structures. Whenever the expansion of a node in the complete clash-free $\mathcal{A}_1$-completion structure has been performed by a sequence of rules captured by a redundant unit completion structure, it is possible to construct a complete clash-free $\mathcal{A}_2$-completion structure by using the simpler non-redundant unit completion structure instead.\qed
\end{proofsketch}

As we still employ the redundancy rule in this version of the algorithm, a complete $\mathcal{A}_2$-completion structure has in the worst case a double exponential number of nodes in the size of the program. As such:

\begin{proposition}\label{prop:complexucp}
$\mathcal{A}_2$ runs in the worst-case in double exponential time.
\end{proposition}

\section{Discussion and Outlook}\label{sec:discussion}

Our optimized algorithm runs in the worst case in double exponential time: this is not a surprise as the scope of the technique introduced here is saving time by avoiding redundant local computations. The worst-case running complexity of the algorithm depends on the depth of the trees which have to be explored in order to ensure completeness of the algorithm and on the fact that anywhere blocking is not feasible. Even with classical subset blocking one has to explore an exponential number of nodes across a branch in order for the algorithm to terminate. Thus, the only factor which would improve the worst-case performance is finding a termination condition which considers nodes in different branches. At the moment this seems highly unattainable. 

The next step of our work is the evaluation of the new algorithm. We expect it will perform considerably better than the original algorithm in returning positive answers to satisfiability checking queries, while it might still take considerable time in the cases where a predicate is not satisfiable. Especially problematic are cases like the one described in Example \ref{ex:blocking} where there exists a unit completion structure which locally satisfies the predicate checked to be satisfiable, but the predicate is actually unsatisfiable. An obvious strategy for implementation is to establish a limit on the depth of the explored structures: in practice it is highly improbable that if there exists a solution, it can be found only in an open answer set of a considerable size: actually, it is quite hard to come up with examples of such situations.

We note that there are also related optimization approaches to ours which do not improve on the worst-case complexity of algorithm, but, which in practice prove to be considerably better than the original algorithms. A knowledge compilation technique for reasoning with the DL $\mathcal{ALC}$ is described in \cite{furbachguentherobermaierFLAIRS09}. First, all sub-concepts of a concept which are conjunctions of simple concepts and role restrictions are computed in the form of so-called \emph{paths} which are sets of simple concepts and role restrictions. Paths which contain contradictory concepts (\emph{links}) are removed, as well as paths which are super-sets of other paths: this is similar to our method in what concerns removing local contradictions and redundancy. However, we also remove redundancies in the set of dependencies between atoms in the model. Then, role restrictions are considered: all links for 'potentially reachable' concepts from the original concept are removed and a so-called \emph{linkless graph} is obtained. Unlike that, we investigate only structures of depth 1: we consider that pre-computing structures with higher depth would be an overkill. The linkless graph is exploited for checking concept consistency and answering subsumption queries. Both reasoning tasks take in the worst case exponential time.

\bibliographystyle{plain}
\bibliography{biblio}

\end{document}